\documentclass[twocolumn]{aastex63}

\newcommand{\kms}{km\,${\rm s}^{-1}$}
\newcommand{\smy}{$M_\odot\,{\rm yr}^{-1}$}

\received{\today}

\submitjournal{ApJ}

\shorttitle{Orbit of WR\,133}
\shortauthors{Richardson et al.}

\graphicspath{{./}{figures/}}

\begin{document}

\title{The First Dynamical Mass Determination of a Nitrogen-rich Wolf-Rayet Star using a Combined Visual and Spectroscopic Orbit}

\correspondingauthor{Noel D. Richardson}
\email{noel.richardson@erau.edu}

\author[0000-0002-2806-9339]{Noel D. Richardson}
\affiliation{Department of Physics and Astronomy, Embry-Riddle Aeronautical University, 3700 Willow Creek Road, Prescott, AZ 86301, USA}

\author{Laura Lee}
\affiliation{Department of Physics and Astronomy, Embry-Riddle Aeronautical University, 3700 Willow Creek Road, Prescott, AZ 86301, USA}

\author[0000-0001-5415-9189]{Gail Schaefer}
\affiliation{The CHARA Array of Georgia State University, Mount Wilson Observatory, Mount Wilson, CA 91023, USA}

\author[0000-0003-0642-8107]{Tomer Shenar}
\affiliation{Institute of Astronomy, KU Leuven, Celestijnenlaan 200D, 3001, Leuven, Belgium}

\author[0000-0002-2090-9751]{Andreas A. C. Sander}
\affiliation{Armagh Observatory, College Hill, Armagh, BT61 9DG, Northern Ireland, UK}

\author{Grant M. Hill}
\affiliation{W. M. Keck Observatory, 65-1120 Mamalahoa Hwy, Kamuela, HI 96743, USA}

\author[0000-0001-7343-1678]{Andrew G. Fullard}
\affiliation{Department of Physics and Astronomy, Michigan State University, 567 Wilson Rd, East Lansing, MI 48824}
\affiliation{Department of Physics and Astronomy, University of Denver, 2112 E. Wesley Ave., 80210, USA}

\author[0000-0002-3380-3307]{John D. Monnier} 
\affiliation{Department of Astronomy, University of Michigan, Ann Arbor, MI 48109, USA}

\author[0000-0002-2208-6541]{Narsireddy Anugu}
\affiliation{Steward Observatory, Department of Astronomy, University of Arizona, Tucson, USA}
\affiliation{School of Physics and Astronomy, University of Exeter,  Exeter, Stocker Road, EX4 4QL, UK}

\author[0000-0001-9764-2357]{Claire L Davies} 
\affiliation{School of Physics and Astronomy, University of Exeter, Exeter, Stocker Road, EX4 4QL, UK}

\author[0000-0002-3003-3183]{Tyler Gardner}
\affiliation{Department of Astronomy, University of Michigan, Ann Arbor, MI 48109, USA}

\author[0000-0001-9745-5834]{Cyprien Lanthermann}
\affiliation{Instituut voor Sterrenkunde, KU Leuven, Celestijnenlaan 200D, 3001 Leuven, Belgium}
\affiliation{Institut de Planetologie et d’Astrophysique de Grenoble, Grenoble 38058, France}

\author[0000-0001-6017-8773]{Stefan Kraus}
\affiliation{School of Physics and Astronomy, University of Exeter, Exeter, Stocker Road, EX4 4QL, UK}

\author[0000-0001-5980-0246]{Benjamin R. Setterholm}
\affiliation{Department of Astronomy, University of Michigan, Ann Arbor, MI 48109, USA}

\begin{abstract}

We present the first visual orbit for the nitrogen-rich Wolf-Rayet binary, \object{WR\,133} (WN5o + O9I) based on observations made with the CHARA Array and the MIRC-X combiner. This orbit represents the first visual orbit for a WN star and only the third Wolf-Rayet star with a visual orbit. The orbit has a period of 112.8 d, a moderate eccentricity of 0.36, and a separation of $a$= 0.79 mas on the sky. We combine the visual orbit with an SB2 orbit and {\it Gaia} parallax to find that the derived masses of the component stars are {$M_{\rm WR}$ = $9.3\pm1.6 M_\odot$ and $M_{\rm O}$ = $22.6\pm 3.2 M_\odot$}, with the large errors owing to the nearly face-on geometry of the system combined with errors in the spectroscopic parameters. We also derive an orbital parallax that is identical to the {\it Gaia}-determined distance. We present a preliminary spectral analysis and atmosphere models of the component stars, and find the mass-loss rate in agreement with polarization variability and our orbit. However, the derived masses are low compared to the spectral types and spectral model. Given the close binary nature, we suspect that WR\,133 should have formed through binary interactions, and represents an ideal target for testing evolutionary models given its membership in the cluster NGC 6871.

\end{abstract}


\section{Introduction} \label{sec:intro}

The most fundamental parameter for a star is its mass, which is {only accurately} measured through application of Kepler's Laws in binary systems. While eclipsing binaries have been the standard for these analyses, some rare types of stars are not yet known to have members in eclipsing systems. In the case of massive stars, it has been shown that most massive stars reside in multiple systems \citep[see, e.g., ][]{2009AJ....137.3358M, 2014ApJS..215...15S, 2015AJ....149...26A}, and it has become an observational fact that binary evolution can dominate the end results of these stars \citep{2012Sci...337..444S, 2013ApJ...764..166D}. In the case of massive stars, the correct interpretation of binary stellar populations could explain the observed SEDs of galaxies throughout history as stripped stars in binary systems changes the ultraviolet flux in a galaxy. However, for this to work properly, we need to have systems for which we can pinpoint stellar parameters and constrain binary evolution \citep{2017PASA...34...58E}.

Some of the best examples of binaries showing evidence of past mass transfer are ones with classical Wolf-Rayet (WR) stars. These stars have lost their hydrogen envelopes through the processes of stellar winds and, possibly, binary interactions.

This analysis focuses on WR\,133, a relatively under-studied binary classified as WN5o+O9I by \citet{1996MNRAS.281..163S}, where the `o' suffix denotes no measurable hydrogen in the WR spectrum and the `N' denotes it as nitrogen-rich. The system is a bright member of NGC 6871 \citep{2015MNRAS.447.2322R}, a cluster residing at a distance of 2.14$\pm$0.07 kpc according to \citet{2009Ap.....52..235M}. The system is used in the calibration of WR star parameters by \citet{2015MNRAS.447.2322R}. The binary orbit of the system was best studied by \citet{1994ApJ...432..770U}, who found a 112.4 d period, a moderate eccentricity of 0.39, and a clear double-lined status. The system shows some polarimetric variability due to changing observed wind geometry throughout the orbit \citep{1989ApJ...347.1034R}, which was recently studied by \citet{2020AJ....159..214F}, who found a lower limit to the inclination to be 115.9$^\circ$. Systems with closer to face-on inclination angles are difficult to constrain using only polarimetric variability. The associated masses for the 115.9$^\circ$ inclination are unrealistically low, meaning that the system would likely be closer to face-on, and perhaps could be resolved with long-baseline interferometry.

In this paper, we present the first long baseline optical interferometry of WR\,133, which spatially resolves the binary and allows us to compute a combined astrometric and spectroscopic orbit. In Section 2, we present our interferometric observations and the spectroscopy used in this analysis. In Section 3, we describe the measurements made and present the orbital elements of the system. We present a combined spectroscopic model in Section 4, and then discuss these results and conclude this study in Section 5.

\section{Observations}
\subsection{Spectroscopy}

\citet{1994ApJ...432..770U} presented the best spectroscopic orbit of the system that has been utilized regularly when describing the system. These 25 observations were taken with the Dominion Astrophysical Observatory 1.83 m telescope (DAO), and had a resolution element of 1.33\AA. In addition, we collected 34 more observations with the same telescope and spectrograph configuration, so that our total dataset spans the years 1986--2013. The observations cover the wavelength range of 5150--6000\AA, allowing for measurement of the He II $\lambda$5411 emission feature for the WR star, and the absorption features of He II $\lambda$5411, He I $\lambda$5876 and O III $5592$ from the O star. A variety of detectors were used given the time-span of the dataset, but all had the same pixel size, thus maintaining the same resolution for the entire data set. The additional emission features from the WR star, C IV $\lambda \lambda$5801,5812 and He I $\lambda$ 5876 are blended with each other and several diffuse interstellar bands or telluric lines and thus are not measured. \citet{1994ApJ...432..770U} also discuss an intermittent problem with the data where the wavelength solution shifts in a zero point, and so the zero point is checked with the deep, sharp interstellar lines from Na D. 

In addition to the DAO data, some additional high-resolution spectra were collected from the ELODIE archive\footnote{http://atlas.obs-hp.fr/elodie/} \citep{2004PASP..116..693M} as well as from the PolarBase\footnote{http://polarbase.irap.omp.eu/} archive \citep{1997MNRAS.291..658D, 2014PASP..126..469P}, which provided some data at much higher spectral resolution. The data from PolarBase were taken with the ESPaDOnS spectropolarimeter on the CFHT, and includes data analyzed by \citet{2014ApJ...781...73D} in a search for magnetic fields. For consistency, we only measured the same lines as available in the majority of our spectra from DAO. 

\begin{figure}
	\includegraphics[angle=90,width=\columnwidth]{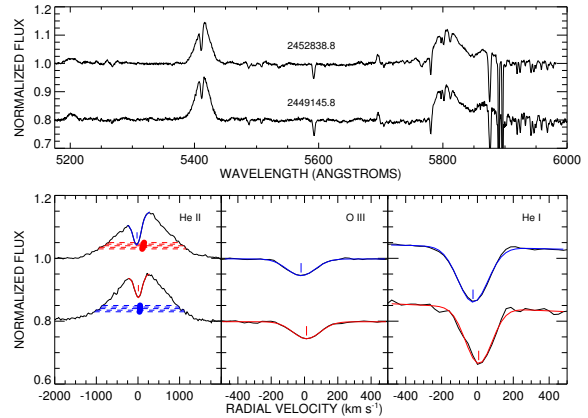}
    \caption{Two example DAO spectra showcasing our measurement routine. The top panel shows examples at different radial velocity extremes. On the bottom panels, we show the red and blue extrema for the Gaussian-fitted O-star lines, and the bisected line levels and measured velocities for the WR star. }
    \label{fig:example_figure}
\end{figure}

\begin{figure*}[ht!]
	\includegraphics[angle=90, width =7in]{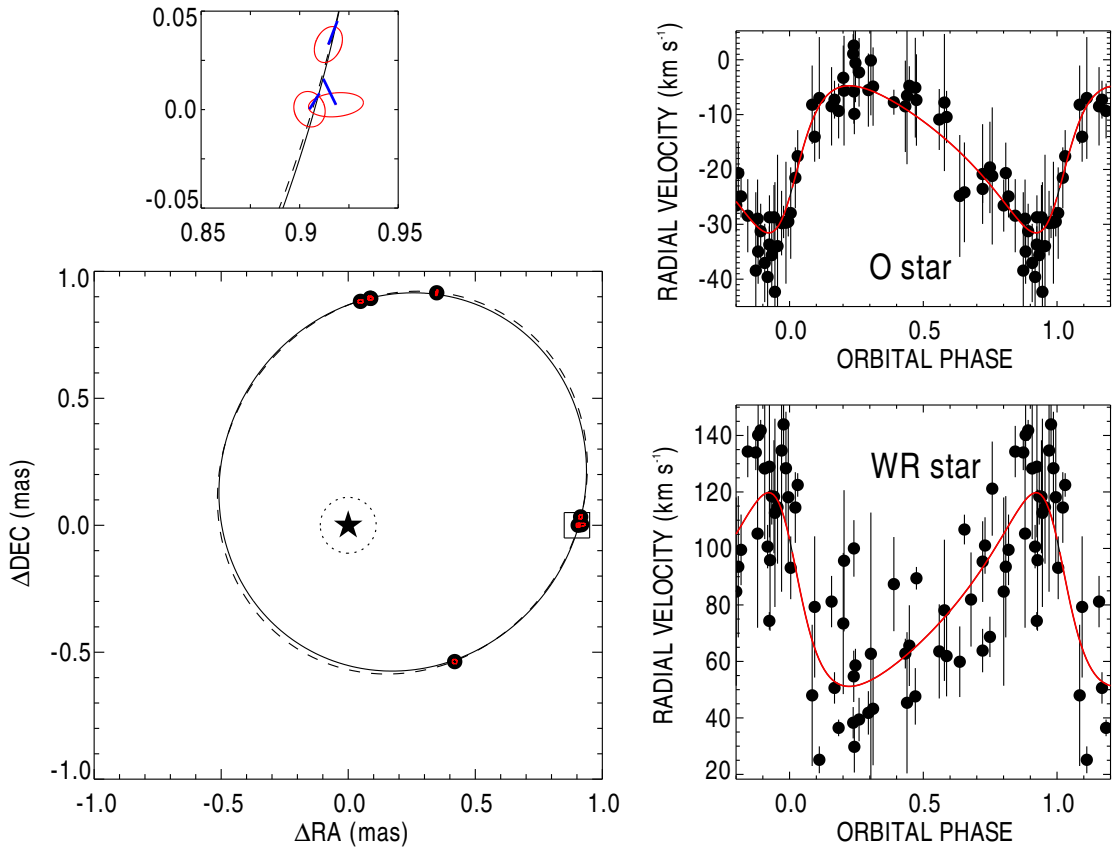}
    \caption{The visual orbit derived for WR\,133 is shown on the left, where we held the distance to be fixed to the {\it Gaia} distance. The central star represents the O star, while the WR star is moving counterclockwise on the orbit. The top panel represents a zoomed-in view of the orbit containing three measurements and highlights the errors of the interferometric measurements, with the connecting blue line showing the offset from the calculated position at that epoch. The solid line shows the fit with the distance held fixed to the {\it Gaia} distance, with the dashed line showing the fit with all parameters free. On the right, we show the spectroscopic orbits of the two component stars. }
    \label{orbit}
\end{figure*}

\subsection{CHARA Array Interferometry}

Interferometric observations were obtained with the CHARA Array, which has six one-meter telescopes in a Y-shaped array with baselines ranging from 34--330 m \citep{2005ApJ...628..453T}
and the upgraded MIRC-X combiner \citep{2004SPIE.5491.1370M,2018SPIE10701E..23K, 2018SPIE10701E..24A, 2020arXiv200712320A} in 2019 and 2020.  The observations were recorded in the PRISM50 mode which provides a spectral resolution of $R = 50$ in the H-band between 1.4 and 1.7 micrometer.  The data were reduced using the MIRC-X data reduction pipeline, version 1.2--1.3\footnote{https://gitlab.chara.gsu.edu/lebouquj/mircx\_pipeline.git.} to produce calibrated visibilities and closure phases.  During the reduction, we applied the bispectrum bias correction included in the pipeline and set the number of coherent coadds to 5. Increasing the number of coherent coadds can improve the signal to noise, but averaging over too long a time in comparison to the atmospheric seeing can bias the calibration. In general, the difference between the binary solutions based on the 5 and 10 coherent coadd reductions was smaller than the formal errors, and the number of coherent coadds of 5 reduction typically gave a lower $\chi^2$ value in the binary fit. We applied minimum uncertainties of 5\% to the visibilities and 0.3 deg to the closure phases based on the typical calibration uncertainties expected for MIRC-X data. 

The calibrators observed on each night are listed in Table \ref{CHARAobs}.  We adopted uniform disk diameters in the H-band for the calibrators from the JMMC stellar diameter catalog \citep{2017yCat.2346....0B}.  We inspected the visibilities and closure phases of the calibrators to check for binarity, and also examined the results from the automated pipeline \citep{2020arXiv200712320A} which calibrates the calibrators against each other and uses CANDID \citep{2015A&A...579A..68G} to determine companion detection limits.  We did not find any reliable binary detections for the calibrators.
Due to the proximity in the sky to WR\,140, we used some of the same calibrator stars as given in the analysis of Thomas et al. (submitted), with the diameter estimates taken from \citet{2017yCat.2346....0B}. All calibrators and data sets are tabulated in Table \ref{CHARAobs} The closure phase calibration is well tested by \citet{2020arXiv200712320A}, so that we are confident that the sign of the closure phases are correct. The reduction pipeline is discussed in detail by \citet{2020arXiv200712320A}, and we refer the reader to that publication for detailed discussion of the reductions.

\begin{table*}
\begin{minipage}{160mm}
\centering
\caption{{Calibrator stars used for the CHARA data}
\label{CHARAobs}}
\begin{tabular}{l c c c c}
\hline \hline
Night   &   $N_{\rm data sets}$ &   Baselines   &   Calibrator  &   $\theta_{\rm UD}$ (mas) \\
\hline
2019-July-01    &   1   &   E1-W2-W1-S2-E2      &   HD 178538   &   0.2487$\pm$0.0062   \\
                &       &                       &   HD 191703   &   0.2185$\pm$0.0055   \\
                &       &                       &   HD 197176   &   0.2415$\pm$0.0058   \\
                &       &                       &   HD 201614   &   0.3174$\pm$0.0074   \\
2019-July-02    &   2   &   E1-W2-W1-S2-S1-E2   &   HD 178538   &   0.2487$\pm$0.0062   \\    
                &       &                       &   HD 191703   &   0.2185$\pm$0.0055   \\
                &       &                       &   HD 197176   &   0.2415$\pm$0.0058   \\
2019-September-04 &   1 &   E1-W2-W1-S2-S1-E2   &   HD 170585   &   0.3576$\pm$0.0087   \\  
                &       &                       &   HD 170977   &   0.3409$\pm$0.0082   \\
                &       &                       &   HD 191703   &   0.2185$\pm$0.0055   \\
                &       &                       &   HD 197176   &   0.2415$\pm$0.0058   \\
                &       &                       &   HD 201614   &   0.3174$\pm$0.0074   \\
2019-September-05 &   2 &   E1-W2-W1-S2-S1-E2   &   HD 170585   &   0.3576$\pm$0.0087   \\  
                &       &                       &   HD 170977   &   0.3409$\pm$0.0082   \\
                &       &                       &   HD 178538   &   0.2487$\pm$0.0062   \\  
                &       &                       &   HD 191703   &   0.2185$\pm$0.0055   \\
                &       &                       &   HD 197176   &   0.2415$\pm$0.0058   \\
                &       &                       &   HD 201614   &   0.3174$\pm$0.0074   \\
2019-September-13 & 1   &   E1-W1-S2-S1-E2      &   HD 191703   &   0.2185$\pm$0.0055   \\
                &       &                       &   HD 197176   &   0.2415$\pm$0.0058   \\
2020-June-23    &   1   &   E1-W2-W1-S2-S1-E2   &   HD 178538   &   0.2487$\pm$0.0062   \\
                &       &                       &   HD 191703   &   0.2185$\pm$0.0055   \\
                &       &                       &   HD 197176   &   0.2415$\pm$0.0058   \\
                &       &                       &   HD 201614   &   0.3174$\pm$0.0074   \\
\hline \hline
\end{tabular}
\end{minipage}
\end{table*}

\section{Measurements and the Orbital Motion}

For the WR star, we measured radial velocities of the He II $\lambda$5411 emission line through bisecting the line (see Fig.~\ref{fig:example_figure}). We measured the bisector at five different heights above the normalized continuum, with levels chosen between 3 and 5\% above the continuum. For each level, the velocity on the red and blue wing was interpolated over a smoothed emission line, and then averaged between the two line wings. The average velocity was then used for the measurement. Errors were estimated and then combined in quadrature for both the standard deviation of these measurements and the differences between the measured velocities of both interstellar Na D absorption lines. We reanalyzed all the data reported by \citet{1994ApJ...432..770U} for consistency of the measurements, and because the former measurements were made with Gaussian fits of the line, which could cause discrepancies between the observations. All velocities are tabulated in an online table.

The O star velocities were measured through Gaussian fits to the He II $\lambda$5411, O III $\lambda$5592, and He I $\lambda$5876 absorption lines. The errors for each individual measurement were estimated through the combination in quadrature of the errors of the centroid position and the error in the wavelength calibration, estimated by the difference in velocity for the fit of the two interstellar Na D absorption lines. The final averaged error was then calculated by combining the standard deviation of these measurements and the maximum individual error of the measurements in quadrature. 

The relative astrometry of the binary was measured by using the calibrated interferometric data with the same approach as \citet{Richardson2016} and Thomas et al.~(submitted). This uses an adaptive grid search to find the best fit binary position and flux ratio using software described by \citet{2016AJ....152..213S}\footnote{This software is available at http://chara.gsu.edu/analysis-software/modeling-software.}. During the binary fit, we fixed the uniform disk diameters based on the Gaia distance and expected radius of the O star. For the O9I star, we adopt $R_* = 22.6 R_\odot$ based on the calibration by \citet{Martins2005}, and adopt the same radius for the WR star. While we have better radii from our models (Section 4), the adopted angular diameters were 0.11 mas, which can be considered point sources with the CHARA Array and have little effect on the derived measurements. 

The uncertainties in the binary fit were derived by adding in quadrature errors computed from four sources: the formal covariance matrix from the binary fit, the variation in parameters when changing the coherent integration time used to reduce the data (5 and 10 coherent coadds), and the variation in parameters when changing the wavelength scale by 0.5\% \citep[the formal uncertainty reported by][for the MIRC-X instrument]{2020arXiv200712320A} as well as the variation in binary parameters when changing the visibility calibration by 5\%. In scaling the uncertainties in the position, we added these values in quadrature for the major axis of the error ellipse ($\sigma_{\rm major}$) and scaled the minor axis ($\sigma_{\rm minor}$) to keep the axis ratio and position angle fixed according to the values derived from the covariance matrix. The results of the astrometric measurements are given in Table~\ref{CHARA measurements}, with significant figures dependent on the individual measurements. In addition to the previously discussed parameters, we include the position angle of the error ellipse ($\sigma_{\rm PA}$) in Table~\ref{CHARA measurements}.

\begin{table*}
\begin{minipage}{160mm}
\centering
\caption{{Astrometric Measurements derived from CHARA data}
\label{CHARA measurements}}
\begin{tabular}{l c c c c c c c}
\hline \hline
UT      &       MJD     &   $\theta$    &   $\rho$ & \multicolumn{3}{c}{Error Ellipse ($\sigma_{\mathrm{PA}}$)}    & Flux Ratio    \\
Date         &        &    ($^\circ$)    &   (mas)    &   major (mas) &   minor (mas) &   PA (mas)  &   $f_{\rm WR} / f_{\rm O}$ \\ \hline

2019 July 01    &   58665.310    &   272.069    &   0.9152    &   0.0048    &   0.0032    &   155.3    &   0.2850$\pm$0.0027 \\
2019 July 02    &   58666.222    &   270.154    &   0.9184    &   0.0070    &   0.0030    &   95.4    &   0.2664$\pm$0.0088 \\
2019 July 02    &   58666.417    &   270.013    &   0.9050    &   0.0046    &   0.0039    &   21.4    &   0.2859$\pm$0.0022 \\
2019 September 04    &   58730.273    &   356.868    &   0.8832    &   0.0052    &   0.0035     &   99.4    &   0.2774$\pm$0.0046 \\
2019 September 05    &   58731.255    &   354.493    &   0.8986    &   0.0050    &   0.0042    &   69.2    &   0.2823$\pm$0.0038 \\
2019 September 05    &   58731.320    &   354.341    &   0.8979    &   0.0049    &   0.0025    &     99.8    &   0.2788$\pm$0.0011 \\
2019 September 13    &   58739.226    &   339.177    &   0.9797    &   0.0058    &   0.0017    &    161.5    &   0.2835$\pm$0.0181 \\
2020 June 23    &   59023.262    &   217.971    &   0.6811    &   0.0050    &   0.0043    &   62.2    &   0.2831$\pm$0.0059 \\
\hline \hline
\end{tabular}
\end{minipage}
\end{table*}

\citet{1994ApJ...432..770U} found that the systemic velocities for the orbit of the Wolf-Rayet and O star components differed by about 100 km s$^{-1}$. In order to minimize the errors present in our fit, we began by fitting the orbit of each component star with the method of \citet{1974PASP...86..455M}. This allowed us to then remove the systemic velocity from the radial velocity measurements we made in order to combine all astrometric and radial velocity measurements into a combined orbital fit. The fitting routine we use was also described by \citet{2016AJ....152..213S}. We found that fitting the orbit resulted in a large reduced $\chi^2$-statistic. We were able to get this to a reasonable value by increasing the errors of the WR radial velocity measurements by a factor of two. The errors of the O-star velocities and the astrometry yield orbital fits with a reduced $\chi^2$-statistic of less than unity when considered as a fit with a visual orbit and an SB1, so our errors in the orbit seem to be dominated by the errors in the WR radial velocities. We also note that \citet{1994ApJ...432..770U} found that the time of periastron passage was different for the different component stars when fitting the orbit. This has been observed in other WR binaries and comes from colliding wind excess emission altering the emission line profiles from the WR star, and could explain our large reduced $\chi^2$-statistic from the combined fit.

\begin{table*}
\begin{minipage}{160mm}
\centering
\caption{{Orbital Elements.}
\label{orbit_elements}}
\begin{tabular}{l c c c }
\hline \hline
Orbital Element & {VB+SB2} &   {At Gaia distance}    & \citet{1994ApJ...432..770U} \\
\hline
\multicolumn{4}{c}{Elements of the System}  \\ \hline 
Period (d)            &  $112.780 \pm 0.036$    &   $112.736 \pm 0.073$     &   $112.4\pm 0.2$     \\
$T$ (periastron, MJD)    &    $58701.58\pm0.38$   &   $58701.58\pm0.56 $     &   $47420.5\pm3.6$ (O-star only)    \\
Eccentricity        &   $0.3558 \pm 0.0050$     &   $0.3646 \pm 0.0103$    &    $0.39 \pm 0.07$   \\
$\omega_{\rm O}$ ($^\circ$)     &   $225.3 \pm 6.1$  &    $238.9 \pm 14.9 $   &   $198.9 \pm 10. $    \\ \hline
\multicolumn{4}{c}{Elements of the Visual Orbit}  \\ \hline 
$a$ (visual, mas)        &   $0.7863 \pm 0.0060$  &   $0.7791\pm 0.0024$     &   $\cdots$    \\
Inclination ($^\circ$)          &   $160.44 \pm 1.86$ &    $ 162.08 \pm 1.74$   &   $\cdots$   \\
$\Omega$ ($^\circ$)            &   $171.5 \pm 6.5$  &    $ 186.1 \pm 15.6$   &   $\cdots$ \\
\hline
\multicolumn{4}{c}{Elements of the Spectroscopic Orbit}  \\ \hline 
$K_{\rm O}$ (km s$^{-1}$)       &   $14.63 \pm 1.51$ &   $13.41 \pm 2.34$      & $16.9\pm 2.1$  \\
$K_{\rm WR}$ (km s$^{-1}$)       &   $32.30 \pm 3.02$ &   $32.56 \pm 3.98$     &  $34.4 \pm 7.4$  \\
$\gamma_{\rm O}$ (km s$^{-1}$)    &   $-15.09 \pm 0.48$  &   $-15.09 \pm 0.48$     &  $-20.9 \pm 0.7$  \\
$\gamma_{\rm WR}$ (km s$^{-1}$)  &   $78.1\pm 3.0$  &   $78.1\pm 3.0$   &  $70.2 \pm 4.6$  \\
\hline 
$\chi^2_{\rm red, fit}$  &   1.37               &    1.07    &  $\cdots$ \\
\hline
\multicolumn{4}{c}{Derived Properties}  \\ \hline 
$a_{\rm WR}$ (AU) &     0.934$\pm$0.122         &   1.021$\pm$0.157    &     $\cdots$      \\
$a_{\rm O}$ (AU) &      0.423$\pm$0.058             &  0.421$\pm$0.083     &  $\cdots$         \\
$a_{\rm WR}$ ($R_\odot$) &     200.7$\pm$26.2        &   219.5$\pm$33.7    &     $\cdots$      \\
$a_{\rm O}$ ($R_\odot$) &      90.9$\pm$12.5             &  90.5$\pm$17.8     &  $\cdots$         \\
Distance (kpc)    &     {1.73$\pm$0.17}    &     {1.85 (fixed)}        &   $\cdots$        \\
$M_{\rm O}$ ($M_\odot$)    &      {$18.1\pm 6.5$}   &    {$22.3\pm9.4$}         &   $\cdots$    \\
$M_{\rm WR}$ ($M_\odot$)     &     {$8.2\pm 2.8$}   &    {$9.2\pm 4.0$}   &   $\cdots$    \\
$R_{\rm roche, WR}$ ($R_\odot$)    &     58.8 & 60.1    &   $\cdots$    \\
$R_{\rm roche, O}$ ($R_\odot$)  &     84.3       &    90.1    &   $\cdots$    \\

\hline
\multicolumn{4}{c}{{Adopting Gaia EDR3 distance of 1.86 $\pm$ 0.08 kpc}}  \\ \hline 
{$M_{\rm O}$ ($M_\odot$)}    &      {$22.5\pm 2.5$}   &    {$22.6\pm3.2$}         &   $\cdots$    \\
{$M_{\rm WR}$ ($M_\odot$)}     &     {$10.2\pm 1.3$}   &    {$9.3\pm 1.6$}   &   $\cdots$    \\
\hline
\end{tabular}
\end{minipage}
\end{table*}

{The orbital elements from the simultaneous fit to the visual orbit and spectroscopic radial velocities are given in Table \ref{orbit_elements} in the column labeled VB+SB2.  These parameters are consistent with the orbital elements of \citet{1994ApJ...432..770U} which are included in the last column of Table \ref{orbit_elements} for reference. With a visual orbit, we can combine the spectroscopic parameters of $a_{\rm WR}\sin i$ and $a_{\rm O}\sin i$ to compare with the angular semi-major axis from interferometry. This yields an orbital parallax yielding a distance of 1.73$\pm0.17$ kpc, which is in agreement with the {\it Gaia} DR2 distance of 1.85$^{+0.16}_{-0.14}$ kpc \citep{2020MNRAS.493.1512R} and the {\it Gaia} early DR3 results which imply the same distance with a smaller uncertainty of 1.86$\pm$0.08 kpc \citep{2020arXiv201201533G}. The dynamical masses computed from $M_{\rm WR}\sin^3 i$, $M_{\rm O}\sin^3 i$, and the inclination are $M_{\rm O} = 18.1 \pm 6.5 M_\odot$ and $M_{\rm WR} = 8.2 \pm 2.8 M_\odot$. These masses have large uncertainties due to the error on the inclination and the nearly face-on system geometry and the uncertainty in the semi-amplitudes. We can reduce the uncertainties in the masses by using the {\it Gaia} EDR3 distance to compute the total mass through Kepler's Third Law and combine with the spectroscopic mass ratio ($K_{\rm O}/K_{\rm WR}$) to get slightly larger masses of $M_{\rm O} = 22.5 \pm 2.5 M_\odot$ and $M_{\rm WR} = 10.2 \pm 1.3 M_\odot$ (last two rows of Table 3). Lastly, in this table, we also include the Roche radii of the component stars from the approximation of \citet{1983ApJ...268..368E}, showing that these stars are not filling their Roche lobes.}

{In Table \ref{orbit_elements} we also present a solution where we constrain the orbital parallax to agree with the Gaia distance of 1.85 kpc. We varied the distance within the 1\,$\sigma$ uncertainties, recomputed the orbit fits, and added the difference in the parameters in quadrature with the formal errors from the fitting routine. The orbital parameters at the Gaia distance are consistent with the fit where the distance is not constrained. We adopt this solution, combined with the uncertainties from the Gaia ERD3 parallax, to produce final masses of $M_{\rm O} = 22.6 \pm 3.2 M_\odot$ and $M_{\rm WR} = 9.3 \pm 1.6 M_\odot$.}

\section{A Spectroscopic Model for the System}

\begin{figure*}
\centering
\includegraphics[width=.84\textwidth]{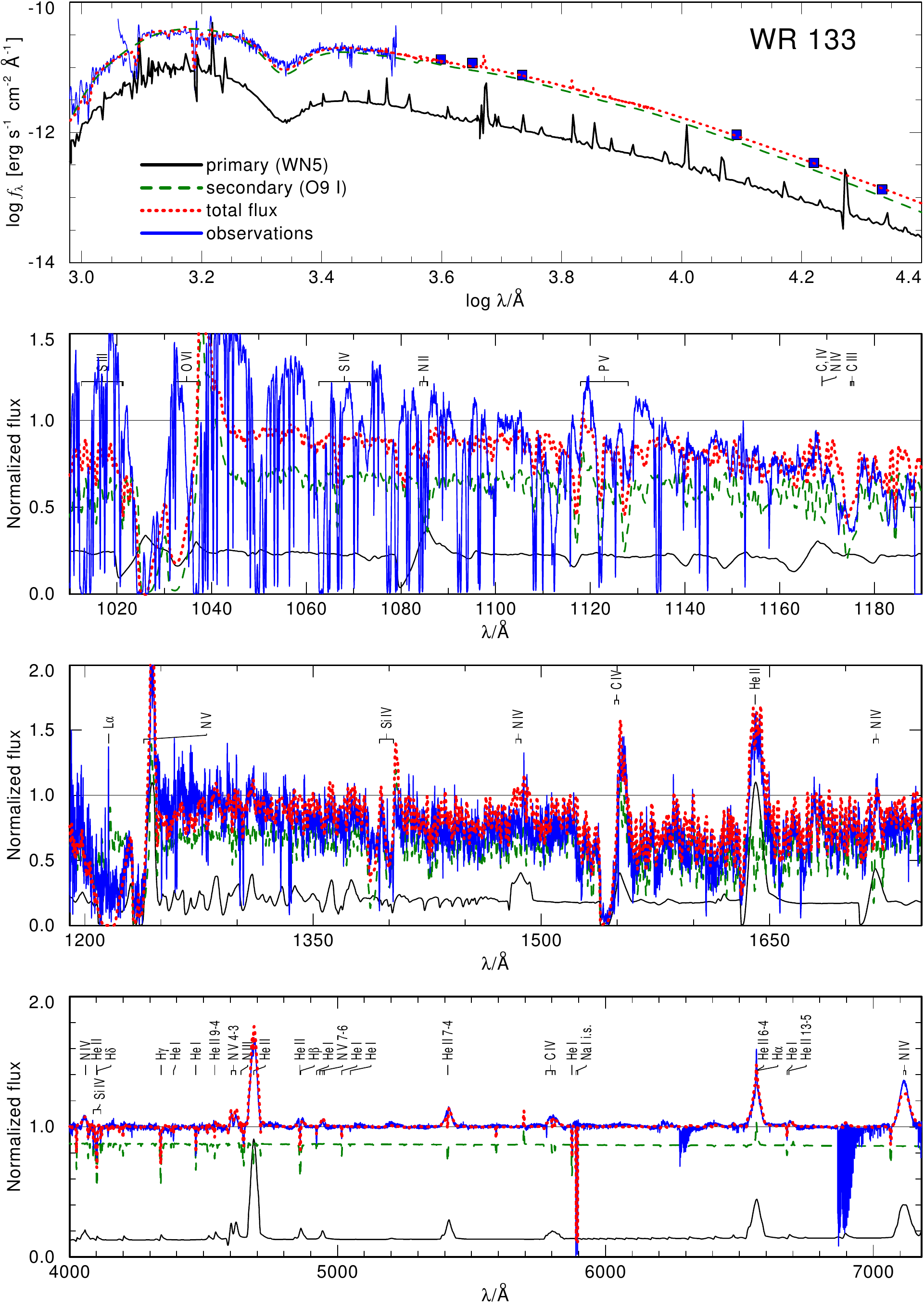}
\caption{{\it Upper panel:} comparison between the observed SED (FUSE, IUE, and UBVJHK photometry, blue lines and squares) and the synthetic PoWR SED modeled for \object{WR~133} (red dotted line). The synthetic SED is the sum of a PoWR model for the WR primary (black solid line) and O-type secondary (green dashed line). {\it Lower panels:} Comparison between the observed and synthetic normalised spectra. See text for details.}
\label{fig:specan}
\end{figure*}

To derive the physical parameters of both stellar components of \object{WR~133}, we utilize the Potsdam Wolf-Rayet (PoWR) model atmosphere code \citep{Hamann2003, Graefener2002, Sander2015} to analyze the available FUSE, IUE, and optical spectra. PoWR is a 1D code that solves the radiative transfer problem in spherical geometry, and is applicable to any hot star with an expanding atmosphere \citep[e.g.,][]{2014A&A...563A..89S, Shenar2015, Ramachandran2019}. The analysis takes advantage of pre-calculated grids for Galactic WN stars \citep{Todt2015} and OB-type stars \citep{Hainich2019}, which enable an estimation of the errors and provide good starting models from which tailored models can be constructed.
The analysis of a binary is possible by combining two PoWR models such that the sum of their spectra reproduce the observations. In the following, we briefly describe the concepts of our PoWR binary analysis. A more detailed description of the methodology is given in \citet{Shenar2016, Shenar2019}.

\begin{table}
\centering
   \caption{Inferred physical parameters of \object{WR~133} from the PoWR spectral analysis}
   \begin{tabular}{lcc} \hline \hline
       Parameter & primary & secondary \\ 
       \hline
       Spectral type    & WN5 & O9 I \\   
       distance\,[kpc]&  \multicolumn{2}{c}{$1.856$ (fixed)}\\          
       flux $f/f_{\rm tot}(H)$ & 0.22 (fixed)  & 0.78 (measured)  \\         
       $T_*$\,[kK]  & $71\pm5$ & $30.0\pm1.0$   \\
       $T_{\rm 2/3}$\,[kK] & $67\pm5$ & $29.5\pm1.0$   \\ 
       $\log g$\,[cgs] & - & $3.30\pm0.15$   \\        
       $\log R_{\rm t}$\,[$R_\odot$] & $0.90\pm0.05$\footnote{{Note that the transformed radius should not be confused with the physical radius of the star, see text for details.}} & -   \\ 
       $\log\,L\,[L_\odot]$  & $5.42\pm0.10$  & $5.29\pm0.10$  \\  
       $R_*\,[R_\odot]$ & $3.4 \pm 0.5$ & $16.6\pm1.0$  \\ 
       $D$ & $4\pm0.3$\,dex  & $20\pm0.3$\,dex  \\ 
       $\beta$ & $1$ (fixed)  & $3\pm1$  \\ 
       $\log \dot{M}$\,[\smy] & $-5.05 \pm 0.15$ & $-6.40 \pm 0.15$  \\
       $v_\infty$\,[\kms{}] & $1600 \pm 100$ & $1800 \pm 200$  \\ 
       $X_{\rm H}$ & $0^{+0.05}_{-0}$ & $0.74\pm0.01 $ \\ 
       $X_{\rm C}/10^{-3}$ & $0.1\pm0.05$ & $2\pm1$  \\ 
       $X_{\rm N}/10^{-3}$ & $15\pm5$ & $0.7\pm0.3$   \\ 
       $X_{\rm O}/10^{-3}$ & - &       $6\pm3$  \\ 
       $v \sin i$\,[\kms{}] & - & $96 \pm 10$  \\
       $v_{\rm mac}$\,[\kms{}] & - & $40 \pm 20$  \\
       $M_{\rm spec}$ & $14 \pm 2$ & $20_{-5}^{+8} $  \\ 
       $E_{B-V}$\,[mag] &  \multicolumn{2}{c}{$0.43 \pm 0.02$}\\              
   \hline
   \end{tabular}
   \label{tab:parameters}
\end{table}

A PoWR model is primarily defined by its temperature $T_*$, gravity $g$, luminosity $L$, mass-loss rate $\dot{M}$, terminal velocity $v_\infty$, and chemical abundances. The temperature $T_*$ refers to the effective temperature at the innermost layer of the model, which is defined at a mean Rosseland optical depth of $\tau_{\rm Ross} = 20$. For OB-type stars, it is almost identical to the effective temperature $T_{\rm 2/3}$, defined at $\tau_{\rm Ross} = 2/3$. For WR stars, the photosphere can be significantly above the hydrostatic layers, which are concealed by the stellar wind. The stellar radius $R_*$ relates to $T_*$ and $L$ via the Stephan-Boltzmann equation $L \propto R_*^2 T_*^4$. 

A helpful parameter used to described PoWR models is the so-called transformed radius $R_{\rm t}$, which characterizes the strength of emission recombination lines: 
\begin{equation}
 R_{\rm {t}} = R_* \left[ \frac{v_\infty}{2500\,{\rm km}\,{\rm s}^{-1}\,}  \middle/  
 \frac{\dot{M} \sqrt{D}}{10^{-4}\,M_\odot\,{\rm yr}^{-1}}  \right]^{2/3}.
\label{eq:Rt}
\end{equation}
Models with given $T_*$ and $R_{\rm t}$ values will tend to exhibit a similar emission spectrum irrespective of $L$, $v_\infty$, and $D$.

The velocity field $v(r)$ in the model smoothly connects the subsonic regime, where the density is determined from quasi-hydrostatic equilibrium \citep{Sander2015}, and the supersonic regime, where the density follows from the wind velocity described by a $\beta-$law \citep{Castor1975}. The winds are assumed to be clumped. Clumping is treated in the microclumping approximation \citep{1998A&A...335.1003H} and is described by the clumping factor $D$, which gives the density ratio between a clumped wind and the equivalent smooth wind. The radial clumping stratification is as described by \citet{Shenar2015} for the O-type component, and is assumed to be constant for the WR component.

The line profiles are characterized by the Doppler velocity $v_{\rm Dop}(r) = \sqrt{v_{\rm th}^2 + \xi^2}$, where $\xi$ is the microturbulent velocity and $v_{\rm th}$ is the thermal velocity \citep{Shenar2015}. The microturbulence is depth-dependent, starting from photospheric values of 100 and 14\,\kms~for the WR and O-type components, respectively \citep{Todt2015, Hainich2019}, and scaling with the wind velocity as $\xi(r) = 0.1 \cdot v(r)$ beyond. 

We include the chemical species H, He, C, N, O, and the iron group elements for both components, and add Mg, Si, P, S for the O-type secondary. The abundances of heavy elements are fixed to solar  \citep{Asplund2009}, but the H, C, N, and O abundances are examined as free parameters in the analysis.

An important quantity in a binary analysis is the flux ratio of both components in a certain band. In principle, it can be estimated from the relative strengths of the O-star features in the spectrum, which would become more or less diluted depending on the adopted light ratio. However, the interferometric measurements described in Section 3 provide us with a direct measure of the light ratio in the $H$-band of $f_{\rm WR} / f_{\rm O}(H) = 0.28$. We fix the light ratio to this measured value.

The effective temperatures $T_*$ are derived by reproducing the ionisation balance of different ionisation stages of a given species, primarily \ion{N}{4} and \ion{N}{5} for the WR primary and \ion{He}{1} and \ion{He}{2} for the O-type secondary. With the fixed light ratio, the wind parameters of the WR primary are determined from the line strength and shape of P-Cygni lines in the UV (most prominently \ion{N}{5} $\lambda \lambda 1239, 1243$ and \ion{C}{4}C\,{\sc iv}\,$\lambda \lambda 1548, 1551$) and the optical recombination lines. The wind parameters of the O-type secondary can also be derived accurately, owing to the presence of significant H$\alpha$ emission, along with  the resonance P-Cygni lines \ion{N}{5} $\lambda \lambda 1239, 1243$, \ion{Si}{4}\,$\lambda \lambda 1394, 1403$, and \ion{C}{4}\,$\lambda \lambda 1548, 1551$. The gravity of the O-type secondary is determined from its Balmer absorption wings. While they are entangled with the WR emission, the gravity can be estimated due to the now fixed WR wind parameters and is found to be typical for O-type supergiants or bright giants \citep[luminosity classes I or II, e.g.,][]{Martins2005}. Lacking photospheric features, the gravity of the WR component cannot be derived. Finally, the abundances are determined from the overall strength of corresponding spectral lines.

The projected rotational velocity $v \sin i$ and macroturbulent velocity $v_{\rm mac}$ of the O-type secondary are determined by utilising the {\sc iacob-broad} tool \citep{Simon-Diaz2007, 2014A&A...562A.135S}, which relies on the Fourier technique and goodness-of-fit analysis. We use the isolated O\,{\sc iii}\,$\lambda 5592$ line for this purpose. The models are convolved with a rotation profile and a radial-tangential profile of the derived Doppler widths \citep{Gray1973}. The values agree with a qualitative inspection between the synthetic and observed line profiles. For the WR primary, $v \sin i$ and $v_{\rm mac}$ cannot be derived.

\begin{figure}
\centering
\includegraphics[width=.5\textwidth]{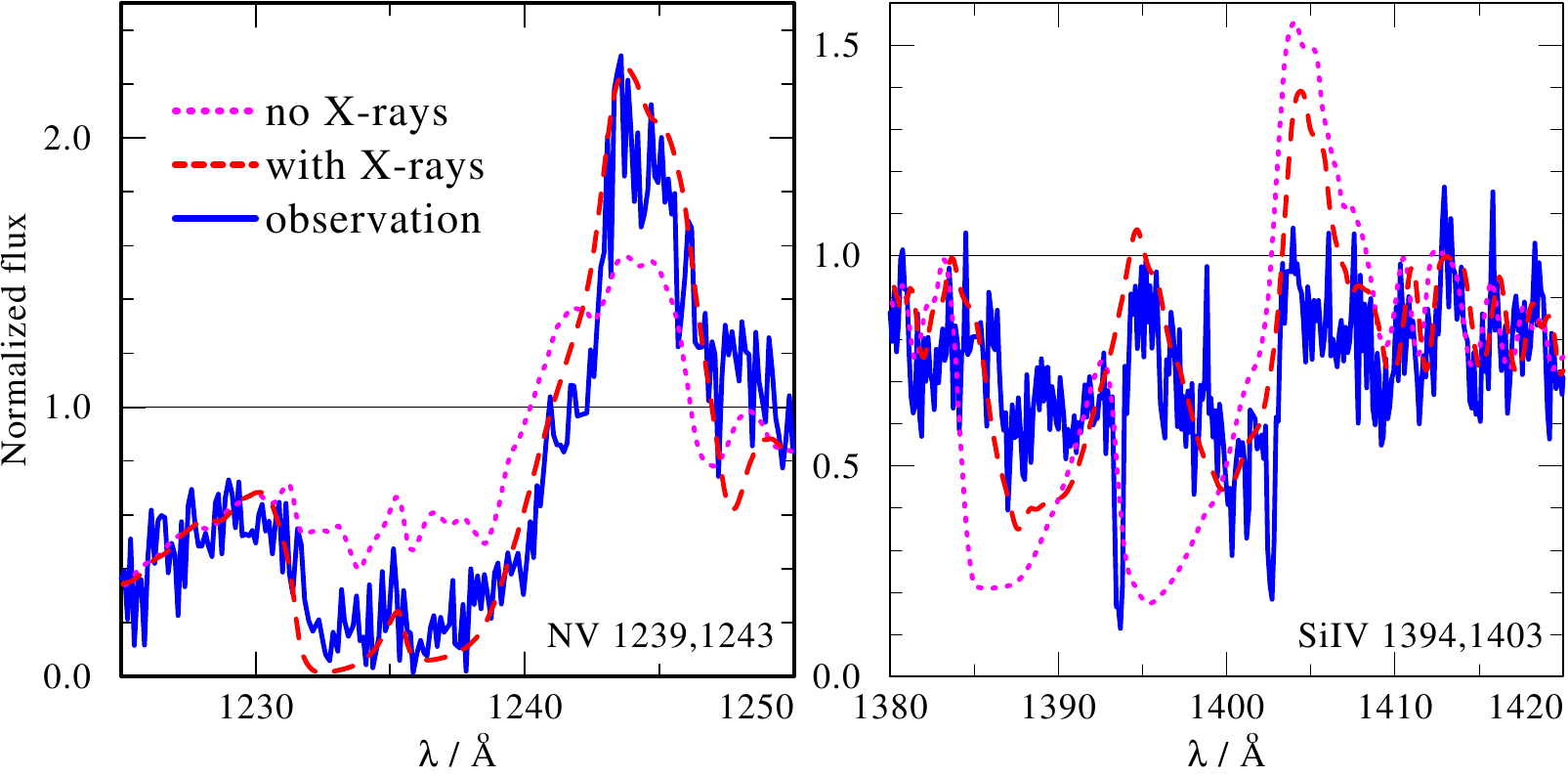}
\caption{Comprison between the observed \ion{N}{5} $\lambda \lambda 1239, 1243$ and \ion{Si}{4}\,$\lambda \lambda 1394, 1403$ IUE spectrum of \object{WR~133} (blue solid line) with the total synthetic PoWR normalised spectrum (sum of O and WR models) with X-rays in the O-type secondary (red dashed line) and without (pink dotted line). Without X-rays in the O model, only the WR star contributes to the \ion{N}{5} $\lambda \lambda 1239, 1243$~doublet, and the \ion{Si}{4}\,$\lambda \lambda 1394, 1403$~doublet is saturated, unlike observed.}
\label{fig:xrays}
\end{figure}

For the WR component, we find no clear indications for $\beta$ values that significantly differ from unity, although we note that there is a degeneracy between $\beta$ and the other parameters. We therefore adopt the typical $\beta = 1$ \citep{Todt2015}. For the O-type component, we find indications for $\beta$ values on the order of 3. Lower $\beta$ values overestimate the width of the H$\alpha$ line, while larger $\beta$ values underestimate it. The terminal velocity is fixed from the P-Cygni lines, and hence $\beta$ is altered to provide a qualitative agreement. The clumping contrasts can only be estimated via electron scattering wings for the WR primary and the relative strength of the P-Cygni and H$\alpha$ line for the O-type secondary. They are found to be typical for the respective spectral types \citep[e.g.,][]{1998A&A...335.1003H, Bouret2012}.

We find very clear indications that X-ray radiation impacts the UV spectrum of the O-type secondary. Through K-Shell transitions (Auger ionization), X-rays can alter the ionization balance in the wind, most prominently populating levels associated with high-ionization lines such as \ion{N}{5} $\lambda \lambda 1239, 1243$ \citep{Cassinelli1979, Oskinova2011}. X-rays are modeled as described in detail in \citet{Baum1992} and \citet{Shenar2015}. In this approach, the X-ray emission is assumed to originate in optically thin filaments of shocked plasma embedded in the
wind, characterized by three parameters: the X-ray temperature $T_{\rm X}$, the filling factor $X_{\rm fill}$, and the onset radius $R_0$. We set $T_{\rm X}$ to $1\,$MK and the onset radius to $R_0 = 2\,R_*$. The filling factor is adjusted such that the observed X-ray luminosity roughly agrees with the observed value of $L_{\rm X} \approx 10^{33}\,{\rm erg}\,{\rm s}^{-1}$ \citep[][]{Pollock1995}. In Fig.\,\ref{fig:xrays}, we show the impact of including X-rays in the PoWR model of the O-type secondary. While the treatment of X-rays is approximate and does not account for non-spherical X-ray irradiation from wind-wind-collisions, it illustrates the necessity of X-rays in reproducing the observed spectrum.

Finally, the reddening and total luminosity of the system are derived by comparing the observed spectral energy distribution (SED) to the sum of the SEDs of both models. The reddening is modeled following \citet{Cardelli1989}. From the total luminosity, the fixed light ratio and the other determined parameters, we then derive the individual luminosities.

The final spectral fit is shown in Fig.\,\ref{fig:specan}. The inferred parameters are given in Table\,\ref{tab:parameters}. The spectroscopic mass for the O-type secondary is derived from $M_{\rm spec} = g\,R_*^2 / G$, where $G$ is the gravitational constant. The spectroscopic mass of the WR primary stems from mass-luminosity relations calculated for chemically homogeneous WR stars by \citet{Graefener2011}, which for a H-free star depends only on $L$. The error on the spectroscopic mass of the WR star is propagated from the error in $L$, 
and is possibly understimated in its lower limit given that the WR primary may be chemically inhomogeneous, which would lower $M$ for a given $L$.

\section{Discussion}

\citet{2012Sci...337..444S} found that about 75\% of massive OB stars should have their evolution impacted by a companion. The general assumption used in the analysis of binary populations is that two stars will interact if the system is born with a semi-major axis of less than $\sim 10$ AU and the star evolves into a red supergiant \citep[e.g., ][]{2017IAUS..329..110S}. During a binary interaction, the system initially becomes more compact until the mass ratio is unity and then increases until mass transfer eventually ceases. This has been seen in an active phase for binaries that could become similar to WR\,133 with systems such as the LBV-like binaries MWC\,314 \citep{2016MNRAS.455..244R} and HDE 326823 \citep{2011AJ....142..201R}. The supergiant companion to the WR star in WR\,133 could have a larger rotation rate than an average field star. The inferred value of $v \sin i$ of 96 km s$^{-1}$ is similar to the average value found for O supergiants by \citet{2014A&A...562A.135S}. However, if there were indeed past binary interactions, the O9I star should be rotating in the orbital plane, and thus would likely be a rapid rotator. {In fact, if we use our derived value for $v \sin i$ and the inclination of the system imply a modestly fast rotation rate of 280 km s$^{-1}$, which is a reasonably fast rotation for a supergiant O star. However, if the O star had a magnetic field from the interaction, then the rotation rate could be drastically less than critical in a short time \citep{2012MNRAS.424.2358P}.}

{\citet{2009MNRAS.400L..20E} used the stellar mass and parameters for the O star in the $\gamma^2$ Velorum system to constrain the age of the $\gamma^2$ Vel stars using both binary as well as single-star evolution models. As the O-star can be rejuvenated during binary interactions, a comparison with single-star models only provides a lower limit for the age of the system. In the case of $\gamma^2\,$Vel, \citet{2009MNRAS.400L..20E} found the system to be about $2\,$Myr older when taking binary evolution into account. Consequently, we can infer a minimum age of our system when we use our measured O star parameters from Table \ref{orbit_elements} as an input for the BONNSAI interface \citep{2014A&A...570A..66S} \footnote{The BONNSAI web-service is available at www.astro.uni-bonn.de/stars/bonnsai.} to explore any evolutionary constraints that might be possible. Given the input parameters from the orbit, along with the model parameters from Table \ref{tab:parameters} (e.g., $\log g$), the star was able to be replicated for the single-star BONNSAI models. With that, we find an initial mass of 27.6$^{+1.9}_{-1.6} M_\odot$, and an age of $5.08 ^{+0.47}_{-0.40}$ Myr. The age for the $\gamma^2$ Velorum system was calculated by \citet{2009MNRAS.400L..20E} to be $5.5 \pm 1$ Myr, in line with the age of the Vela OB association. This means that the WR\,133 system is about as old or even older than $\gamma^2\,$Vel. Moreover, the mass exchange and envelope stripping of the WR component in WR\,133 must have been considerably weaker than in $\gamma^2\,$Vel. }

{It seems that WR\,133 is similar to the $\gamma^2$ Velorum system in that it has had past interactions, and the orbit now has a small eccentricity and a period of $\sim100$ d. The main difference in these systems is likely the progenitor masses. In the case of WR\,133, we see a less massive system than in the case of $\gamma^2$ Vel, as well as nitrogen-rich WR star instead of a C-rich WR star. This means that either WR\,133 was not massive enough to undergo the same evolution as $\gamma^2$ Vel, or that it is not as far along in its evolutionary path. In the case of $\gamma^2$ Vel, we may expect that the larger initial masses would lead to a larger radiative force to drive the stellar winds and evolve to the carbon-rich WR we observe today. }

The component masses we derive for WR\,133 are lower than those inferred with the PoWR model described in Section 4. We note that the large errors in the masses are currently dominated by the seemingly small error of the orbital inclination of 1.85$^\circ$, but due to the nearly face-on system geometry, it propagates to large errors, especially when we also consider the errors in the spectroscopic semi-amplitudes. While our orbital mass of 22.3 $M_\odot$ for the O star agrees with the PoWR model results (20 $M_\odot$), it is noteworthy that \citet{Martins2005} predicts a higher mass of 32 $M_\odot$ for the O9I spectral type.
If the spectral type was O9III instead of O9I, the expected mass from \citet{Martins2005} would be 21$M_\odot$, which is more in line with our findings. We note that many of the spectral classification lines are likely contaminated by the presence of the WR star, potentially blurring the stellar classification of the component stars in the WR\,133 system.

\citet{2020AJ....159..214F} recently published models of the polarimetric variability of WR\,133. With our better-constrained orbital parameters, we can then use the same models for the system to constrain the mass-loss rate of the WR star. Using the parameters from our orbit, such as $\Omega$, the flux ratio, $a_{\rm system}$, and the inclination, the polarimetric model is best fit with a mass-loss rate of the WR star of $(4.0 \pm 0.6)\times 10^{-6} M_\odot {\rm yr}^{-1}$, which is about a factor of 2 lower than the PoWR-derived mass-loss rate. 

The orbit of WR\,133 offers us the chance to better understand the nature of these interactions. The system resides in the Galactic open cluster NGC 6871 \citep{2015MNRAS.447.2322R}. This cluster has a line of sight velocity of $-11$ km s$^{-1}$ \citep{2005A&A...438.1163K}, comparable to the value we derive for $\gamma_{\rm O} = -15.3$ km s$^{-1}$. The cluster has an age of $\approx 10\,$Myr \citep{2005A&A...438.1163K}, meaning that the system had to form and interact in that relatively short time frame.

Finally, we note that there are three well-established masses for WR stars, where the mass comes from a combination of a visual (interferometric) orbit and a double-lined spectroscopic orbit. The two additional systems are $\gamma^2$ Velorum \citep{2017MNRAS.468.2655L} which contains a WC8 star with a mass of 8.9 $M_\odot$, and WR\,140 (Thomas et al., submitted) which contains a WC7 star with a mass of 10.3 $M_\odot$. It is interesting to note that the two WC stars appear to have similar masses to the WN star in WR\,133. However, we continue to be in the range of very small number statistics, and any differences in these masses could be reflecting different conditions for binary interactions in the past.

WR\,133 provides a rare chance to constrain stellar wind structure models, in particular for the spectroscopically unconstrained masses of WN stars, which is a key ingredient for our perception of WR-type mass loss \citep{2020MNRAS.491.4406S}.
For these purposes, an improvement of our orbital solution will be vital.
This can be done with improved spectroscopic measurements to both constrain the spectral types and better measure the radial velocities. We will need more measurements with the CHARA Array to improve our visual orbit and better constrain the inclination. The two proposed orbits in this paper differ by about 0.02 mas near periastron, which should be possible to measure with the MIRC-X combiner in the future. It also allows us to use polarimetry to provide independent constraints on the mass-loss rate of the WR star, in turn providing for more calibrated measurements for many WR$+$O binaries in the future.

\acknowledgments

{We thank our anonymous referee for an insightful report that led to better clarity in this paper.} This work is based in part upon observations obtained with the Georgia State University Center for High Angular Resolution Astronomy Array at Mount Wilson Observatory.  The CHARA Array is supported by the National Science Foundation under Grant No. AST-1636624 and AST-1715788.  Institutional support has been provided from the GSU College of Arts and Sciences and the GSU Office of the Vice President for Research and Economic Development. Some of the time at the CHARA Array was granted through the NOAO community access program (NOAO PropID: 17B-0088; PI: Richardson). MIRC-X received funding from the European Research Council (ERC) under the European Union's Horizon 2020 research and innovation programme (grant No.\ 639889). This work is also based in part on spectral data retrieved from the ELODIE archive at Observatoire de Haute-Provence (OHP). This research has made use of the Jean-Marie Mariotti Center \texttt{searchcal} and \textit{aspro2} service.

Some of LL's participation was funded by the Embry-Riddle Aeronautical University's Undergraduate Research Institute.
T.S. acknowledges funding from the European Research Council (ERC) under the European Union’s Horizon 2020 research and innovation programme (grant agreement number 772225).
A.A.C.S. is supported by STFC funding under grant number ST/R000565/1.
J.D.M. acknowledges support from NASA-XRP NNX16AD43G and NSF-AST 1909165.
S.K. acknowledges support from ERC Starting Grant "ImagePlanetFormDiscs" (Grant Agreement Number 639889).

\vspace{5mm}
\facilities{CHARA, DAO:1.83m}

\clearpage

\bibliography{WR133}{}
\bibliographystyle{aasjournal}

\end{document}